\documentclass[letterpaper,pop,showpacs,preprint,superscriptaddress,english]{revtex4-1}
\usepackage{lmodern}

\usepackage[T1]{fontenc}
\usepackage[latin9]{inputenc}
\usepackage{amsmath}
\usepackage{amssymb}
\usepackage{graphicx}
\usepackage{esint}

\makeatletter

\newenvironment{lyxlist}[1]
{\begin{list}{}
{\settowidth{\labelwidth}{#1}
 \setlength{\leftmargin}{\labelwidth}
 \addtolength{\leftmargin}{\labelsep}
 }}
{\end{list}}

\usepackage{amsthm}\usepackage{latexsym}\usepackage{bm}\usepackage{amsfonts}

\makeatother

\begin{document}

\title{Spectral and structural stability properties of charged particle
dynamics in coupled lattices }

\author{Hong Qin}

\affiliation{Plasma Physics Laboratory, Princeton University, Princeton, NJ 08543}

\affiliation{Department of Modern Physics, University of Science and Technology
of China, Hefei, Anhui 230026, China}

\author{Moses Chung}

\affiliation{Department of Physics, Ulsan National Institute of Science and Technology,
Ulsan 689-798, Korea}

\author{Ronald C. Davidson}

\affiliation{Plasma Physics Laboratory, Princeton University, Princeton, NJ 08543}

\author{Joshua W. Burby}

\affiliation{Plasma Physics Laboratory, Princeton University, Princeton, NJ 08543}
\begin{abstract}
It has been realized in recent years that coupled focusing lattices
in accelerators and storage rings have significant advantages over
conventional uncoupled focusing lattices, especially for high-intensity
charged particle beams. A theoretical framework and associated tools
for analyzing the spectral and structural stability properties of
coupled lattices are formulated in this paper, based on the recently
developed generalized Courant-Snyder theory for coupled lattices.
It is shown that for periodic coupled lattices that are spectrally
and structurally stable, the matrix envelope equation must admit matched
solutions. Using the technique of normal form and pre-Iwasawa decomposition,
a new method is developed to replace the (inefficient) shooting method
for finding matched solutions for the matrix envelope equation. Stability
properties of a continuously rotating quadrupole lattice are investigated.
The Krein collision process for destabilization of the lattice is
demonstrated.
\end{abstract}


\maketitle

\section{Introduction\label{sec:Introduction}}

The transverse focusing lattice is one of the few crucial subsystems
in modern accelerators and storage rings. Most contemporary accelerators
and storage rings are designed based on an uncoupled linear transverse
lattice \cite{Courant58}, where the two degrees of freedom in the
transverse directions are decoupled. Well-known analyses of the effects
of weak coupling on stability properties have left the incorrect impression
that the coupling between the $x$-dynamics and $y$-dynamics always
results in instabilities or other deleterious effects. It has been
realized recently that coupled lattices are not necessarily more unstable
than uncoupled lattices. On the contrary, it is believed that coupled
lattice can be more advantageous in comparison with conventional uncoupled
lattices, especially for high-intensity charged particle beams \cite{Gluckstern79,Roberson1985,Chernin1988,Petillo1989,Krall1995,Smith1998,Petillo2002,Talman95,Barnard96,Kishek99}.
This is because the parameter space for coupled lattices is much larger
than that for uncoupled lattices, and one can explore the larger parameter
space for a coupled lattice to optimize the lattice design.

Of course, the most important consideration in lattice design is its
stability properties. A thorough study of lattice stability requires
one to distinguish two types of stability properties, spectral stability
and structural stability \cite{Krein50,Gelfand55,Moser58,Yakubovich75,DragtBook,Qin14-044001}.
The spectral stability of a linear periodic lattice is determined
by the eigenvalues of the one-period map $M$ of the lattice. If there
exists a vector $v$ such that $M^{l}v\rightarrow\infty$ as $l\rightarrow\infty$,
then the map is spectrally unstable. Otherwise, it is spectrally stable.
This is the most familiar stability property that is often analyzed.
The structural stability of the lattice refers to the robustness of
the spectral stability property of the lattice with respect to a structural
perturbation, such as imperfections in the magnets, or misalignment
of the beam-line. A lattice is structurally unstable if there exists
a spectrally unstable lattice infinitesimally close-by. Otherwise,
it is structurally stable. 

Unfortunately, our understanding of the stability properties of coupled
lattices is far from comprehensive due to the lack of an effective
mathematical tool to describe the coupled dynamics. For 1D uncoupled
dynamics, the \textit{de facto} standard for parameterizing the focusing
lattice is the Courant-Snyder (CS) theory \cite{Courant58}, which
is mathematically elegant and directly linked to the physics of the
beam dynamics. For coupled lattices, several parameterization schemes
have been developed \cite{Teng71,Edwards73,Teng03,Ripken70,Wiedemann07-614,Lebedev10,Dattoli92,Dattoli92b,Dattoli92c}.
But none of these schemes is as effective for coupled lattices as
the CS theory is for uncoupled lattices. Recently, we have developed
a generalized Courant-Snyder theory for coupled lattices \cite{Qin09-NA,Qin09PoP-NA,Qin09-PRL,Chung10,Qin11-056708,Qin13PRL2,Qin13PRL,Chung13,Qin14-044001},
which generalizes every important aspect of the original CS theory
to higher dimensions. Especially, the key components of the original
CS theory, i.e., the envelope function (or the $\beta$ function)
and the associated envelope equation are generalized into a matrix
envelope function and the associated matrix envelope equation. 

In the present study, we apply the generalized CS theory to investigate
the stability properties of coupled lattices. We prove an important
proposition that a necessary condition for a periodic coupled lattice
to be spectrally and structurally stable is that the generalized matrix
envelope equation admits a matched solution. We also show how to apply
the techniques of pre-Iwasawa decomposition \cite{Iwasawa49,deGosson06-42}
and normal form to construct a matched solution of the matrix envelope
equation by simply solving the initial value problem once. This new
method is of great value even for uncoupled lattices. Previously,
one used the conventional shooting method to solve the initial value
problem many times to search for a matched solution. Using the example
of a continuously rotating quadrupole lattice, we illustrate in this
paper how the lattice becomes spectrally unstable through an interesting
process called the Krein collision. 

The paper is organized as follows. In Sec.\,\ref{sec: Hamiltonian},
we describe the spectral and structural stability properties of a
generic linear periodic Hamiltonian system and the associated Krein
collision. The generalized Courant-Snyder theory for coupled lattices
is introduced in Sec.\,\ref{sec:GCS}, and the connection between
stability properties and matched lattice functions are discussed in
Sec\,\ref{sec:Stability-Analysis}. The formalism developed is applied
to study the stability properties of a continuously rotating quadrupole
lattice in Sec.\,\ref{sec:crq}.

\section{Spectral and structural stability properties of linear Hamiltonian
systems\label{sec: Hamiltonian}}

The dynamics of a charged particle in a coupled or uncoupled periodic
focusing system is completely specified by the one-period map. Because
of the Hamiltonian nature of the dynamics, the one-period map is symplectic.
For the linear focusing lattices considered in the present study,
the one-period map is specified by a symplectic matrix $M$. In this
section, we discuss the stability properties of $M$ as a general
symplectic matrix. The calculation and stability analysis of $M$
for a specific choice of focusing lattice will be discussed in Secs.\,\ref{sec:GCS}
and \ref{sec:Stability-Analysis} using the generalized Courant-Snyder
theory. 

\begin{figure}[ptb]
\begin{centering}
\includegraphics[width=3.5in]{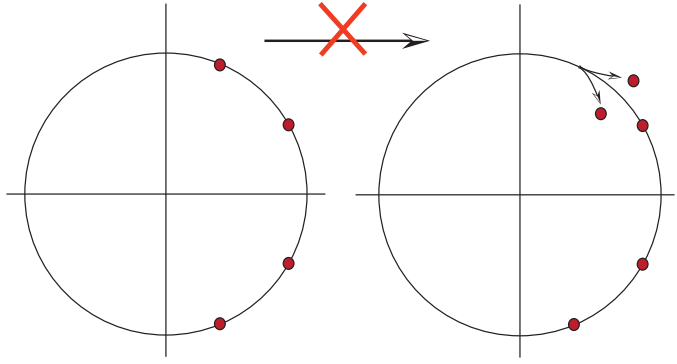} 
\par\end{centering}

\protect\caption{Eigenvalues are forbidden to move off the unit circle for Case 1.
Illustrated here is the case of $n=2.$}

\label{forbidden} 
\end{figure}

\begin{figure}[ptb]
\begin{centering}
\includegraphics[width=3.5in]{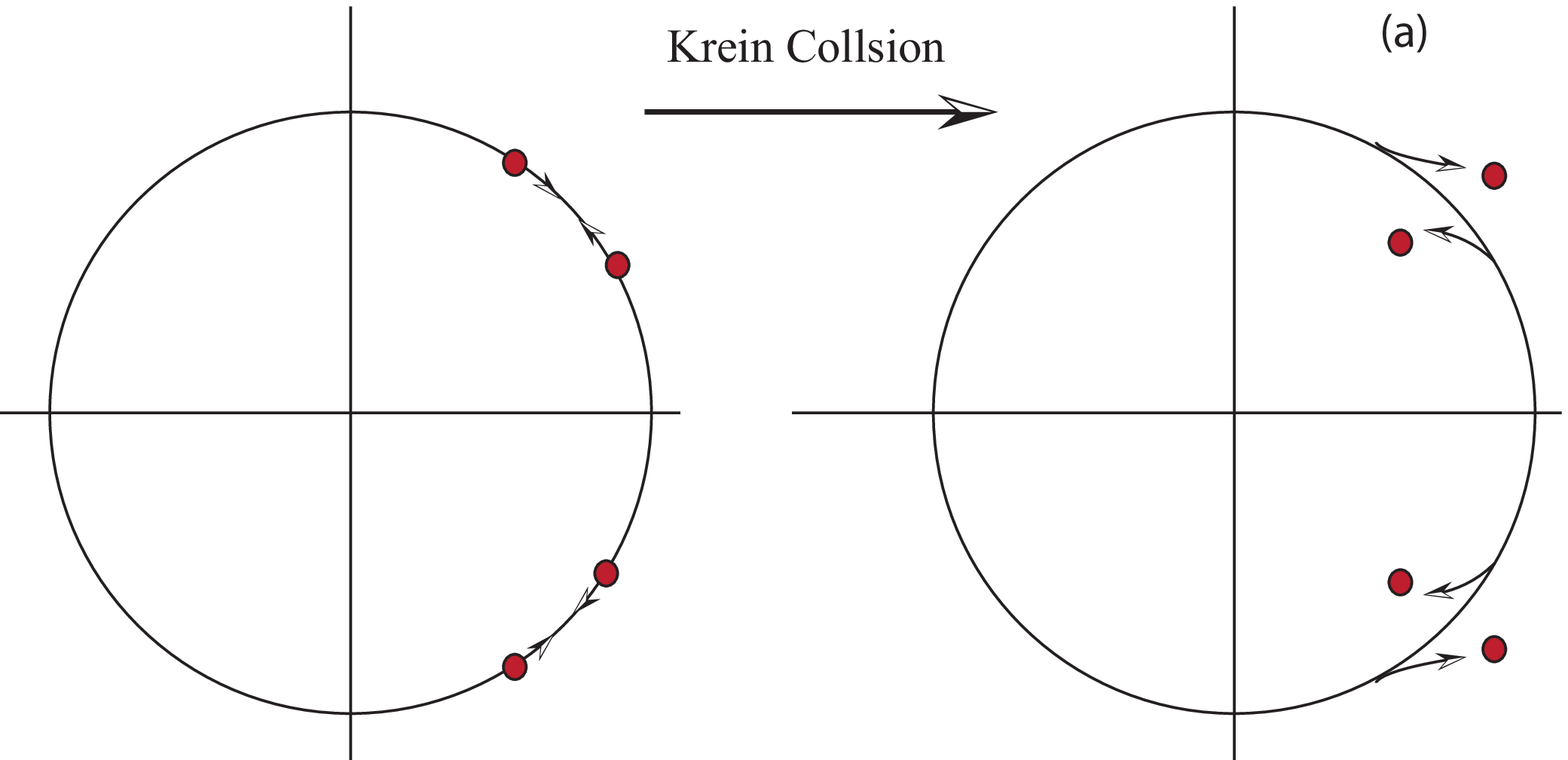} 
\par\end{centering}

\begin{centering}
\includegraphics[width=3.5in]{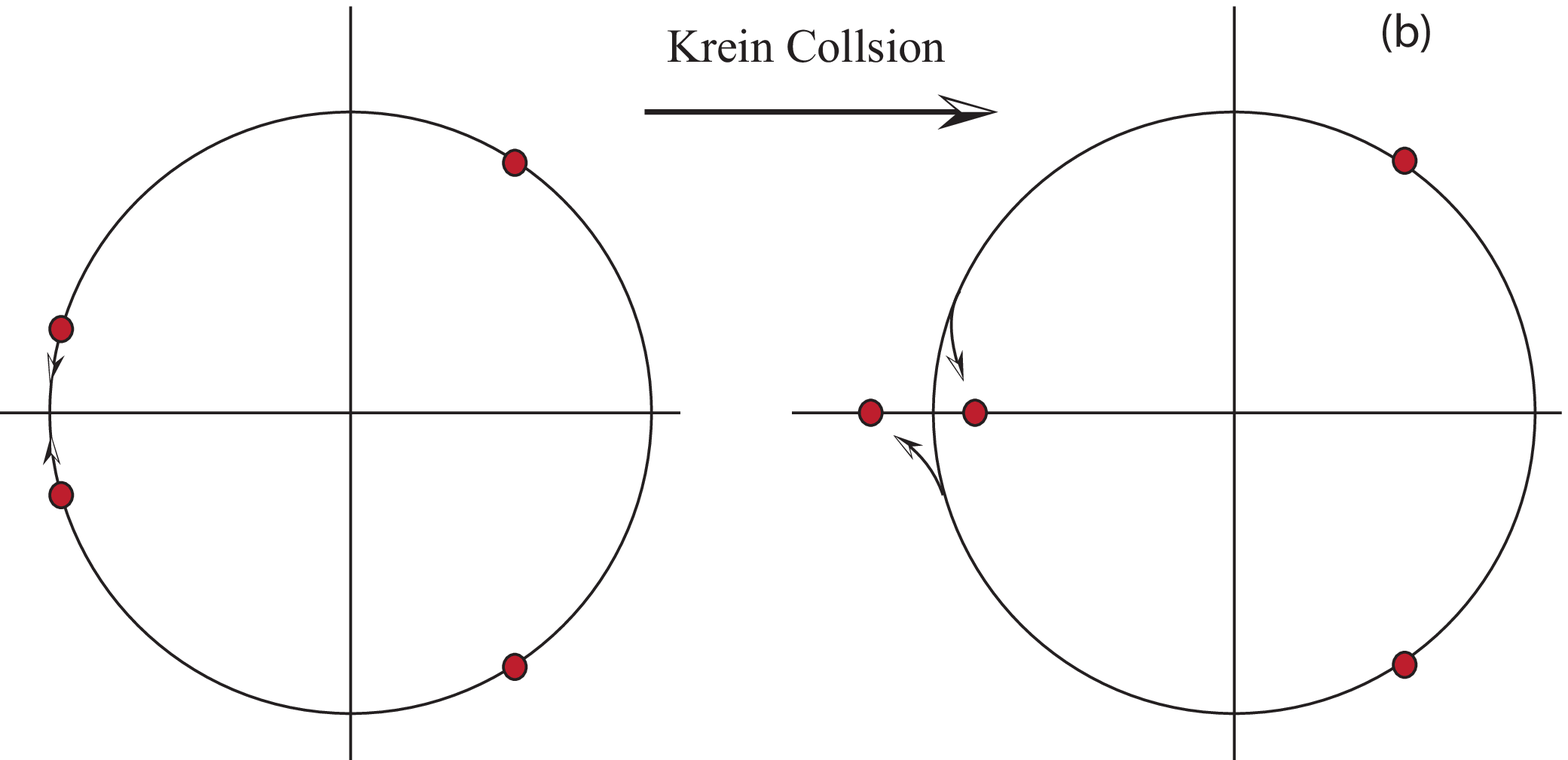} 
\par\end{centering}

\protect\caption{Eigenvalues are allowed to move off the unit circle for Case 2 when
there are repeated eigenvalues. This is the so-called Krein collision.
Illustrated in (a) and (b) are two possibilities for the case of $n=2.$
For each possibility, the eigenvalue distribution on the left is before
the collision, and the one on the right is after the collision.}

\label{Krein} 
\end{figure}

Let the dimension of $M$ be $2n\times2n$, and let $\lambda_{l}\thinspace\thinspace(l=1,...,2n)$
be the eigenvalues of $M.$ It is straightforward to prove that if
$\lambda$ is an eigenvalue of a symplectic matrix, then its inverse
$1/\lambda$ and its complex conjugate $\bar{\lambda}$ are also eigenvalues.
Then, the eigenvalue distribution can be divided into four categories:
\begin{description}
\item [{\textmd{(1)}}] All eigenvalues are distinct and on the unit circle
of the complex plane, i.e., $\left|\lambda_{l}\right|=1$ and $\lambda_{l}\neq\lambda_{m}$
for $l\neq m.$
\item [{\textmd{(2)}}] All eigenvalues are on the unit circle. There are
repeated eigenvalues. But the geometric multiplicity for all eigenvalues
are the same as the algebraic multiplicity, i.e., $Mul_{g}(\lambda_{l})=Mul_{a}(\lambda_{l})$
for all $l.$
\item [{\textmd{(3)}}] All eigenvalues are on the unit circle. There are
repeated eigenvalues with algebraic multiplicity greater than geometric
multiplicity, i.e., $Mul_{g}(\lambda_{l})<Mul_{a}(\lambda_{l})$ for
some $l.$
\item [{\textmd{(4)}}] There exits at least one eigenvalue not on the unit
circle, i.e., $\left|\lambda_{l}\right|\neq1$ for some $l.$
\end{description}
Here, an eigenvalue$\lambda_{l}$ of $M$ is a root of the characteristic
polynomial $Det(I-\lambda M)$, and the algebraic multiplicity of
an eigenvalue $Mul_{a}(\lambda_{l})$ is the order of the root. The
geometric multiplicity of an eigenvalue $Mul_{g}(\lambda_{l})$ is
the number of independent eigenvectors corresponding to the eigenvalue.
In general, $Mul_{g}(\lambda_{l})\le Mul_{a}(\lambda_{l})$. According
to the basic theory of linear algebra, Cases (1) and (2) are spectrally
stable, and Cases (3) and (4) are spectrally unstable. For Cases (1)
and (2), we would like to know whether their spectral stability will
be sustained under a small structural perturbation. Case (1) can also
be shown to be structurally stable by considering the symplectic nature
of $M.$ As the structure of the system is perturbed, the eigenvalues
will move accordingly. However, they cannot move off the unit circle
due to a small structural perturbation on $M$ for Case (1). This
is because for every eigenvalue $\lambda$ off the circle, there exits
another eigenvaule $1/\bar{\lambda}$, which is on the opposite side
of the unit circle. If one of the eigenvalues of Case (1) were allowed
to move off the unit circle, then there would be more than $2n$ eigenvalues.
This forbidden situation is illustrated in Fig.\,\ref{forbidden}
for $n=2$. When there are repeated eigenvalues for Case (2), the
constraints on the locations of the eigenvalues do not prohibit the
eigenvalues moving off the unit circle due to structural perturbations,
which is the so-called Krein collision \cite{Krein50,Gelfand55,Moser58,Yakubovich75,DragtBook,Qin14-044001},
as illustrated in Fig. \ref{Krein} for $n=2.$ Krein collisions preserve
the symmetry of the eigenvalue distribution with respect to the real
axis and the unit circle, and are the only possible pathways in parameter
space for a spectrally stable system to become spectrally unstable.
When this happens, the system is structurally unstable. What is more
interesting is that not all possibilities in Case (2) are structurally
unstable. Case (2) needs to be further divided into two sub-categories:
\begin{lyxlist}{00.00.0000}
\item [{(2.1)}] For all repeated eigenvalues, the corresponding eigenvectors
have the same signatures. 
\item [{(2.2)}] There is at least one repeated eigenvalue whose eigenvectos
have different signatures.
\end{lyxlist}
Here, the signature of an eigenvector $\psi$ of $M$ is defined to
be the sign of its self-product $\left\langle \psi,\psi\right\rangle =\psi^{*}iJ\psi.$
Here, $J$ is the $2n\times2n$ unit symplectic matrix, i.e., 
\begin{equation}
J=\left(\begin{array}{cc}
0 & I_{n}\\
-I_{n} & 0
\end{array}\right).\mbox{}
\end{equation}
The product between two vectors $\psi$ and $\phi$ in general is
defined to be 
\begin{equation}
\left\langle \psi,\phi\right\rangle \equiv\psi^{*}iJ\phi,\label{eq:product}
\end{equation}
where $\psi^{*}$ is the complex conjugate of $\psi^{T}.$ Krein \cite{Krein50},
Gel'fand and Lidskii \cite{Gelfand55}, and Moser \cite{Moser58}
proved that Case (2.1) is structurally stable, and that Case (2.2)
is structurally unstable. This is the celebrated Krein-Gel'fand-Lidskii-Moser
theorem. 


Let's use the example of a 1D uncoupled lattice ($n=1$) to demonstrate
the process of a Krein collision. For a 1D uncoupled periodic lattice,
the one-period transfer map $M$ is \cite{Qin14-044001}
\begin{align}
 & M=S_{0}^{-1}PS_{0},\\
 & P=\left(\begin{array}{cc}
\cos\phi & \sin\phi\\
-\sin\phi & \cos\phi
\end{array}\right),\\
 & S_{0}=\left(\begin{array}{cc}
w_{0}^{-1} & 0\\
-\dot{w}_{0} & w_{0}
\end{array}\right),\\
 & \phi=\int_{0}^{T}\dfrac{dt}{\beta\left(t\right)},\,\\
 & \beta\left(t\right)=w^{2}\left(t\right)\,,
\end{align}
where $w(t)$ is a matched envelope function, $w_{0}=w(0)$ and $\dot{w}_{0}=\dot{w}(0)$
are the initial conditions for $w(t)$, and $\phi$ is the one-period
phase advance. The envelope function $w(t)$ is determined by the
envelope equation
\begin{equation}
\ddot{w}+\kappa_{q}\left(t\right)w=w^{-3}\,.\label{w1}
\end{equation}
Since $M$ is similar to $P,$ and the eigenvalues and signatures
are preserved by a similarity transformation {[}see Eq.\,\eqref{eq:spsiphi}{]},
the spectral and structural stability properties of the lattice is
completely determined by the phase advance matrix $P$. The eigenvalues,
eigenvectors, and signatures are 
\begin{align}
\lambda_{+} & =\cos\phi+i\sin\phi\,,\,\,\,\psi_{+}=(1,i)^{T}\,,\,\,\,\sigma_{+}=-1\,,\\
\lambda_{-} & =\cos\phi-i\sin\phi\,,\,\,\,\psi_{-}=(1,-i)^{T}\,,\,\,\,\sigma_{-}=1\,.
\end{align}
Evidently, the system is spectrally and structurally stable when $\phi\neq n\pi,$
which corresponds to Case (1). As the system parameters varies, the
phase advance $\phi$ changes. A Krein collision occurs at $\phi=n\pi,$
where $\lambda_{+}=\lambda_{-}$ and $\sigma_{+}=-\sigma_{-}$. This
is Case (2.2), which is structurally unstable. Starting from a stable
lattice with a small phase advance, we can increase the focusing strength
and thus the phase advance gradually. The system is stable until the
phase advance approaches $\pi.$

\section{Generalized Courant-Snyder theory \label{sec:GCS}}

To prepare for the investigation of the stability properties for general
coupled lattice, we briefly summarize here the generalized Courant-Snyder
theory, a thorough description of which can be found in Ref. \cite{Qin14-044001}.

The linear dynamics of a charged particle relative to the fiducial
orbit are governed by a general time-dependent Hamiltonian \cite{Micheloetti95-166}
of the form 
\begin{align}
H & =\frac{1}{2}z^{T}\Lambda z\,,\,\,\,\Lambda=\left(\begin{array}{cc}
\kappa\left(t\right) & R\left(t\right)\\
R\left(t\right)^{T} & m^{-1}\left(t\right)
\end{array}\right)\,.\label{H}
\end{align}
Here, $z=\left(x,y,p_{x},p_{y}\right)^{T}$ are the transverse phase
space coordinates, and $\kappa(t),$ $R\left(t\right)$ and $m^{-1}\left(t\right)$
are time-dependent $2\times2$ matrices. The matrices $\kappa(t)$,
$m^{-1}\left(t\right)$, and $\Lambda$ are also symmetric. In this
general Hamiltonian, the quadrupole component is in the diagonal terms
of the $\kappa\left(t\right)$ matrix. The off-diagonal terms of $\kappa\left(t\right)$
contain the skew-quadrupole and dipole components. The solenoidal
component and the torsion of the fiducial orbit \cite{Hoffstaetter95}
are included in the $R\left(t\right)$ matrix. The symplectic matrix
specifying the map between $z_{0}$ and $z=M(t)z_{0}$ is 
\begin{align}
 & M(t)=S^{-1}P^{T}S_{0},\label{eq:M}\\
 & S=\left(\begin{array}{cc}
w^{-T} & 0\\
(wR-\dot{w})m & w
\end{array}\right),
\end{align}
where subscript ``0'' denotes initial conditions at $t=0$, and
$w$ is a $2\times2$ envelope matrix function satisfying the matrix
envelope equation 
\begin{equation}
\frac{d}{dt}\left(\frac{dw}{dt}m-wRm\right)+\frac{dw}{dt}mR^{T}+w\left(\kappa-RmR^{T}\right)-\left(w^{T}wmw^{T}\right)^{-1}=0\,.\label{w}
\end{equation}
In Eq.\,\eqref{eq:M}, $P\in Sp(4)\bigcap SO(4)=U(2)$ is a symplectic
rotation, which is the generalized phase advance, determined by
\begin{align}
 & \dot{P}=-P\left(\begin{array}{cc}
0 & -\mu\\
\mu & 0
\end{array}\right)\,,\\
 & \mu\equiv\left(wmw^{T}\right)^{-1}.
\end{align}

Alternatively and preferably, the transfer matrix $M(t)$ can be expressed
in terms of a symmetric envelope matrix $u(t),$ which is defined
to be the symmetric component of $w(t)$ in its polar decomposition,
\begin{align}
 & u(t)\equiv\sqrt{\beta(t)}\thinspace,\\
 & \beta(t)\equiv w^{T}(t)w(t)\thinspace.
\end{align}
In terms of $u,$ the transfer matrix is
\begin{align}
 & M(t)=S_{u}^{-1}P_{u}^{-1}S_{u0}\,\label{eq:Mu}\\
 & S_{u}\equiv\left(\begin{array}{cc}
u^{-1} & 0\\
(uR-Du-\dot{u})m & u
\end{array}\right)\,\\
 & D\equiv L_{umu}^{-1}\left[(um\dot{u}-\dot{u}mu)+u(Rm-mR^{T})u\right]\,,\label{eq:D}
\end{align}
where $P_{u}\in Sp(4)\bigcap SO(4)=U(2)$ is a symplectic rotation
determined by the differential equation 
\begin{align}
\dot{P}_{u} & =-P_{u}\left[\left(\begin{array}{cc}
0 & \mu_{u}\\
-\mu_{u} & 0
\end{array}\right)-\left(\begin{array}{cc}
D & 0\\
0 & D
\end{array}\right)\right]\,,\label{eq:Pu}\\
\mu_{u} & \equiv\left(umu^{T}\right)^{-1}\,.
\end{align}
In Eq.\,\eqref{eq:D}, $L_{V}^{-1}$ is the inverse of Lyapunov operator
defined as
\begin{equation}
L_{V}(X)=VX+XV\mbox{}\,,\label{eq:Ly}
\end{equation}
for a symmetric, positive-definite matrix $V$. A detailed discussion
of the Lyapunov operator can be found in Ref. \cite{Qin14-044001}.
The envelope matrix $u(t)\equiv\sqrt{\beta(t)}$ is determined from
the differential equation for $\beta,$ 
\begin{align}
\ddot{\beta} & =2e-fg-g^{T}f^{T}-\beta h-h^{T}\beta+2m^{-1}\beta^{-1}m^{-1}\label{eq:beta}\\
e & \equiv\dot{u}Du+\dot{u}^{2}-uD^{2}u-uD\dot{u}\,\\
f & \equiv uDu+u\dot{u}\,,\\
g & \equiv(\dot{m}-Rm+mR^{T})m^{-1}\,,\\
h & \equiv(\kappa-RmR^{T}-\dot{R}m-R\dot{m})m^{-1}\,,\\
\dot{u} & =L_{\sqrt{\beta}}^{-1}(\dot{\beta})\,.
\end{align}
Equation \eqref{eq:beta} is a second-order ordinary differential
equation for $\beta$, since every term on the right-hand-side is
a function of $\beta$ and $\dot{\beta}$. 

For every $t$, $M(t)$ is specified by two $n\times n$ symmetric
matrices $\beta$ and $\dot{\beta},$ and a $U(n)$ matrix $P_{u}.$
The dimension of $M(t)$ is thus $(n^{2}+n)/2+(n^{2}+n)/2+n^{2}=n(2n+1)$,
as expected for symplectic matrices. Another important advantage of
using the symmetric envelope matrix $u$ over the unsymmetric envelope
matrix $w$ is that Eq.\,\eqref{eq:Mu} enables the application of
advanced techniques of pre-Iwasawa decomposition and normal form to
find a matched solution for the $\beta$ matrix without using the
(inefficient) shooting method.

\section{Stability analysis and matched lattice functions\label{sec:Stability-Analysis}}

For practical applications of coupled lattices, it is desirable to
design a coupled lattice belongs to Cases (1) and (2.1), which are
both spectrally and structurally stable. As mentioned previously,
the parameter space satisfying this condition for a coupled lattice
is larger than that for an uncoupled lattice. The generalized Courant-Snyder
theory described in Sec.\,\ref{sec:GCS} provides an effective tool
to study the stability properties of coupled lattices. One of the
important result is that if a matched solution for $\beta$ exists,
then the stability property of a general coupled lattice is completely
determined by the phase advance matrix $P_{u}$. Using a matched solution
for $\beta,$ the one-period map is
\begin{equation}
M(T)=S_{u0}^{-1}P_{u}(T)^{T}S_{u0}\,,
\end{equation}
which indicates that $M(T)$ is similar to the inverse of the phase
advance $P_{u}(T)^{T}$, and thus has the same eigenvalues and multiplicity
as $P_{u}(T)^{T}$. Because $P_{u}(T)^{T}$ is a rotation, its eigenvalues
are on the unit circle. Now we show that the phase advance $P_{u}(T)$
also determines the structural stability of $M(T)$. For an eigenvector
$\psi$ of $M(T)$, $S_{u0}\psi$ is an eigenvector of $P_{u}(T)^{T},$
and the product between the two eignervectors defined in Eq.\,\eqref{eq:product}
is preserved by the similarity transformation, i.e., 
\begin{equation}
\left\langle S_{u0}\psi,S_{u0}\phi\right\rangle =\psi^{*}S_{u0}^{T}iJS_{u0}\phi=\psi^{*}iJ\phi=\left\langle \psi,\phi\right\rangle \,.\label{eq:spsiphi}
\end{equation}
Therefore, $P_{u}(T)^{T}$ and $M(T)$ have the same eigenvalues,
signatures, and thus structural stability properties.

As in the case of a 1D uncoupled lattice, matched solutions for $\beta$
are much more preferable than unmatched solutions, because the lattice
functions can be completely determined by a matched $\beta$ solution
in one lattice period, i.e., the lattice functions are periodic in
terms of the lattice period. On the other hand, if an unmatched $\beta$
solution is used, the $\beta$ function and other lattice functions
have to be solved in the entire time domain of $0<t<\infty.$ Does
a matched $\beta$ solution always exist? The answer is negative.
What are the conditions for the existence of a match $\beta$ solution?
We prove now that for Cases (1) and (2.1), Eq.\,\eqref{eq:beta}
admits matched solution for $\beta.$ The proof utilizes the technique
of normal form and pre-Iwasawa decomposition \cite{Iwasawa49,deGosson06-42}
for symplectic matrices.

First, let's invoke the established result that for Case (1), $M$
can always be transformed into the following normal form with a symplectic
matrix $A,$
\begin{align}
 & M=ANA^{-1},\label{eq:NF}\\
 & N=\left(\begin{array}{cccc}
R_{1}\\
 & R_{2}\\
 &  & ...\\
 &  &  & R_{n}
\end{array}\right),\\
 & R_{l}=\left(\begin{array}{cc}
\cos\phi_{l} & \sin\phi_{l}\\
-\sin\phi_{l} & \cos\phi_{l}
\end{array}\right).
\end{align}
Obviously, $N\in Sp(2n)\bigcap SO(2n)=U(n)$ is a symplectic rotation.
This fact is proved \cite{DragtBook} from the existence of a complete
set of $2n$ orthonormal eigenvectors $(\psi_{l},\psi_{-l}),\;(l=1,2,...,n)$,
satisfying 
\begin{align}
 & \left\langle \psi_{l},\psi_{m}\right\rangle =\delta_{lm},\label{eq:ylm}\\
 & \left\langle \psi_{-l},\psi_{-m}\right\rangle =-\delta_{lm},\label{eq:y-lm}\\
 & \left\langle \psi_{l},\psi_{-m}\right\rangle =0.\label{eq:yl-m}
\end{align}
Here, $\psi_{l}$ and $\psi_{-l}=\bar{\psi}_{l}$ are a pair of eigenvectors
corresponding to the eigenvalues $\lambda_{l}$ and $\lambda_{-l}=\bar{\lambda}_{l}$,
respectively. Equations \eqref{eq:ylm} and \eqref{eq:y-lm} state
that $\psi_{l}$ and $\psi_{-l}$ have different signatures. The normal
form is actually explicitly constructed. The transfer matrix $A$
is given as \cite{DragtBook}
\begin{equation}
A=\sqrt{2}(\xi_{1},\eta_{1},\xi_{2},\eta_{2},...,\xi_{n},\eta_{n}),
\end{equation}
where $\xi_{l}$ and $\eta_{l}$ are real and imaginary components
of the eigenvector $\psi_{l}$, i.e., $\psi_{l}=\xi_{l}+i\eta_{l}.$ 

We now show that for Case (2.1), such a set of orthonormal bases exists
as well. For a repeated eigenvector $\lambda$ with $Mul_{g}(\lambda)=Mul_{a}(\lambda)$,
the corresponding eigenvectors span a subspace $M_{\lambda}$ of $R^{2n}.$
Because the signature never vanishes in $M_{\lambda},$ we can always
select a set of orthonormal bases for $M_{\lambda}$ through a Gram-Schmidt
process. The subspace $M_{-\lambda}$, a complex conjugate image of
$M_{\lambda}$, has the same structure except that the signature has
the opposite sign. Therefore, for both Case (1) and Case (2.1), the
normal form given by Eq.\,\eqref{eq:NF} exists. 

Second, we apply the pre-Iwasawa decomposition to the symplectic matrix
$A.$ According to the theory of Iwasawa decomposition \cite{Iwasawa49,deGosson06-42},
a symplectic matrix $G$ can always be uniquely factored as 
\begin{equation}
G=P\left(\begin{array}{cc}
Y & 0\\
QY & Y^{-1}
\end{array}\right)\,,
\end{equation}
where $P\in Sp(2n)\bigcap SO(2n)=U(n)$ is a symplectic rotation,
and $Y$ and $Q$ are symmetric. The statement is true as well if
the decomposition is defined alternatively to be
\begin{equation}
G=\left(\begin{array}{cc}
Y & 0\\
QY & Y^{-1}
\end{array}\right)P\,.
\end{equation}

Let the unique pre-Iwasawa decomposition of $A$ be 
\begin{align}
 & A=P_{A}S_{A}\thinspace,\\
 & S_{A}=\left(\begin{array}{cc}
Y & 0\\
QY & Y^{-1}
\end{array}\right)_{A}.
\end{align}
Then, the transfer matrix is $M=S_{A}^{-1}P_{A}^{-1}NP_{A}S_{A}.$
We choose the initial conditions for $\beta$ and $\dot{\beta}$ such
that 
\begin{equation}
S_{u0}=S_{A},\label{eq:Su0}
\end{equation}
and the solution of $\beta$ will give the same transfer map
\begin{equation}
M=S_{u}^{-1}P_{u}^{-1}S_{A}=S_{A}^{-1}P_{A}^{-1}NP_{A}S_{A}.
\end{equation}
Thus
\begin{equation}
S_{u}^{-1}P_{u}^{-1}=S_{A}^{-1}P_{A}^{-1}NP_{A}.
\end{equation}
The uniqueness of pre-Iwasawa decomposition requires that
\begin{align}
 & S_{u}^{-1}=S_{A}^{-1},\label{eq:su-1}\\
 & P_{u}^{-1}=P_{A}^{-1}NP_{A}.
\end{align}
Equations \eqref{eq:Su0} and \eqref{eq:su-1} prove that $S_{u}^{-1}=S_{u0}^{-1}$,
i.e., the $\beta$ solution is matched. Thus we have proven the proposition
that for Cases (1) and (2.1), Eq.\,\eqref{eq:beta} admits matched
solutions. This proof is a constructive one, which can be used as
an effective method to find a matched solution for $\beta$. The conventional
method for finding matched solutions is the shooting method, where
one takes an initial condition for $\beta$ and solves the ordinary
differential equation once in one period. In general, the solution
is not matched, i.e., $\beta(T)\neq\beta(0)$ or $\dot{\beta}(T)\neq\dot{\beta}(0).$
The shooting method requires one to estimate a better initial condition
based on the size of the mismatch, and solve the differential equation
again. This iteration is carried out many times until a matched solution
is found. The new method suggested by the above constructive proof
of the existence of matched solution only requires solving Eq.\,\eqref{eq:beta}
once with an arbitrary initial condition. We can construct the one-period
map $M(T)$ using any matched or unmatched solution of Eq.\,\eqref{eq:beta},
then the eigenvectors of $M(T)$ can be calculated. When the set of
bases satisfying Eqs.\,\eqref{eq:ylm}- \eqref{eq:yl-m} exists,
the initial condition for a matched solution is uniquely given by
Eq.\,\eqref{eq:Su0}. This new method applies to both 1D uncoupled
lattices and coupled lattices in higher dimensions. Of course, this
procedure fails when the set of bases satisfying Eqs.\,\eqref{eq:ylm}-
\eqref{eq:yl-m} does not exist. However, for the desirable Cases
(1) and (2.1), such a set of bases exists. Another practical implication
of the existence of matched $\beta$ solution is that when an matched
solution for $\beta$ cannot be found, the lattice must be unstable.

\section{continuously rotating quadrupole lattices\label{sec:crq}}

\begin{figure}[ptb]
\begin{centering}
\includegraphics[width=3.5in]{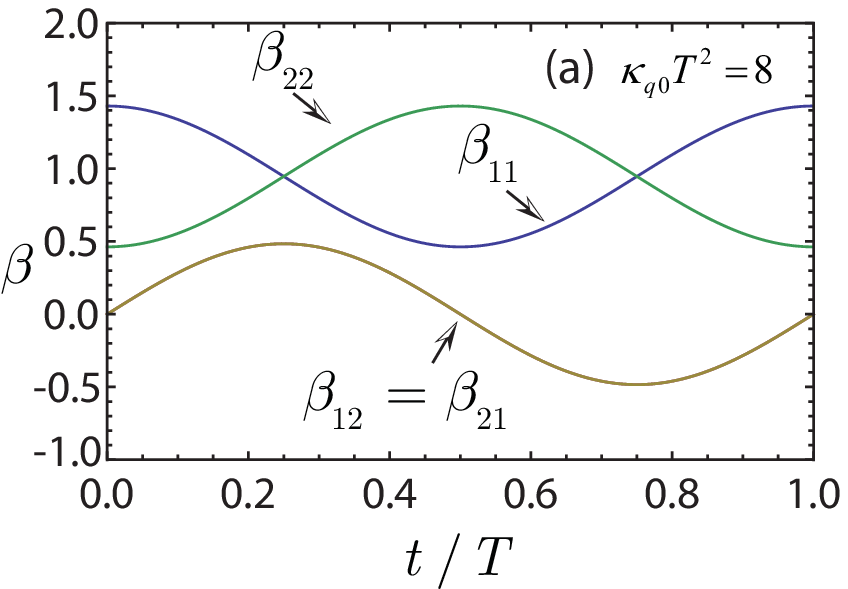} \includegraphics[width=3.5in]{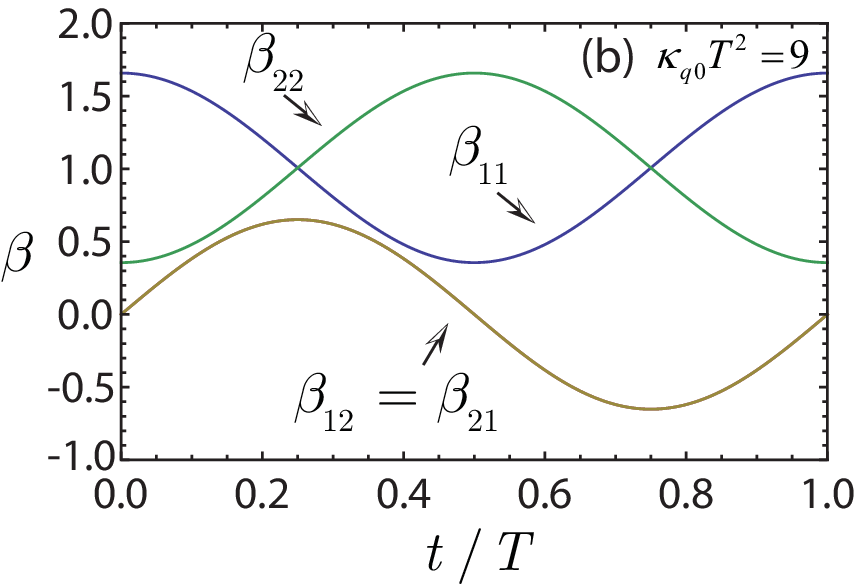} 
\par\end{centering}

\protect\caption{Matched solutions for the cases of (a) $\kappa_{q0}T^{2}=8$ and (b)$\kappa_{q0}T^{2}=9$. }

\label{ms} 
\end{figure}

\begin{figure}[ptb]
\begin{centering}
\includegraphics[width=2.8in]{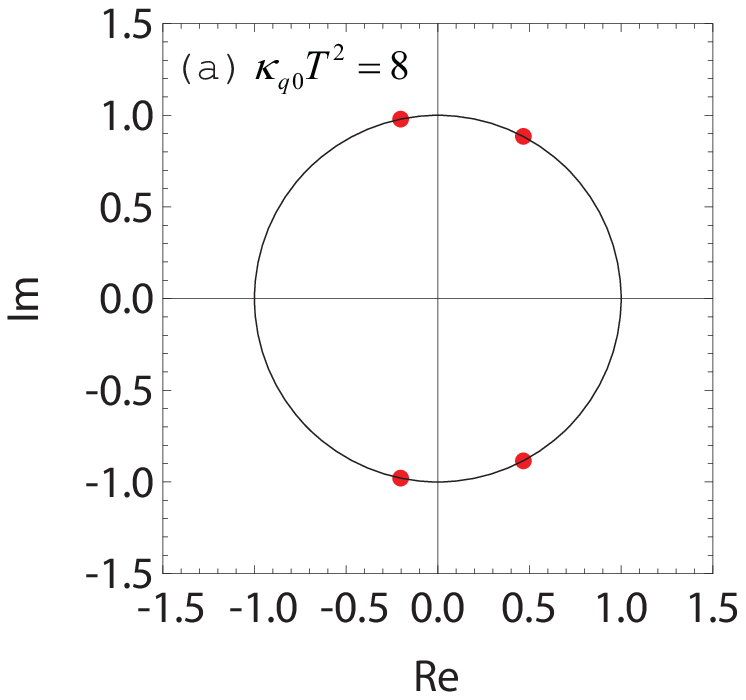} \includegraphics[width=2.8in]{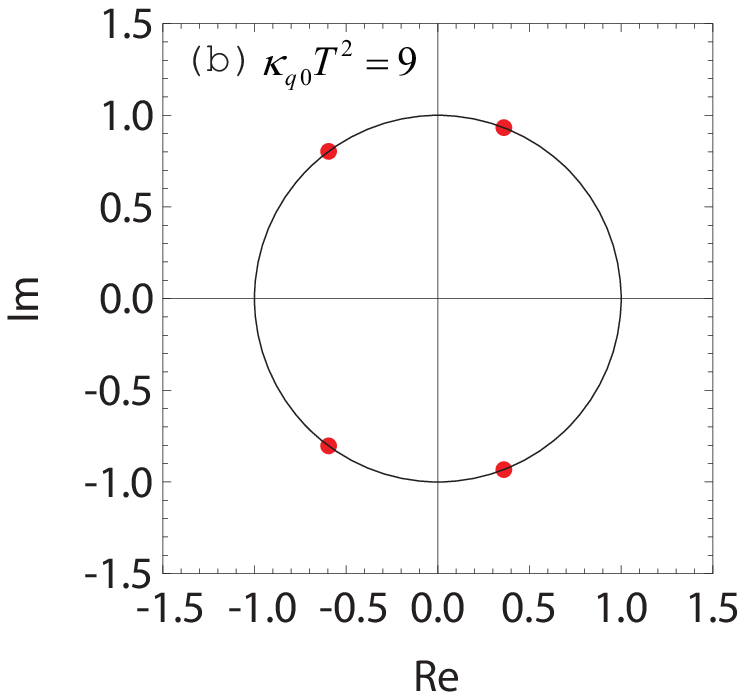}
\includegraphics[width=2.8in]{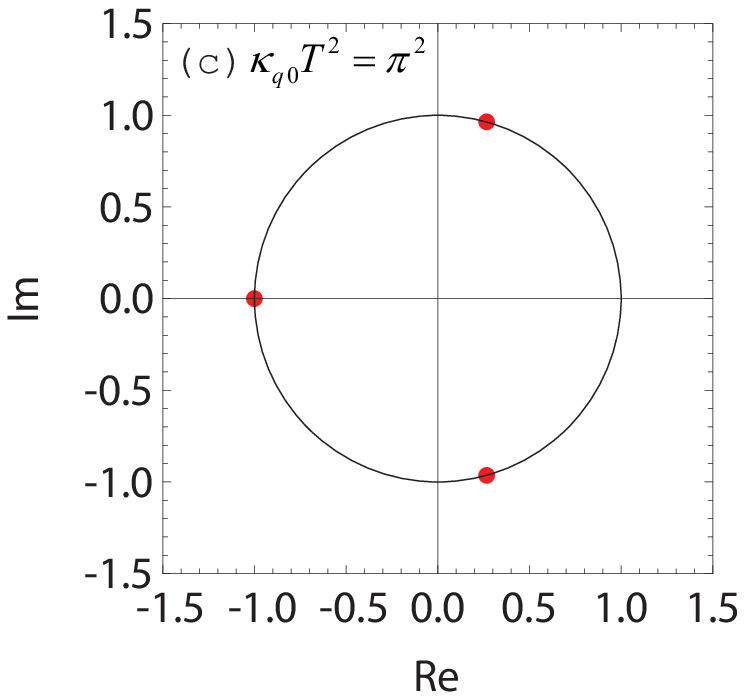} \includegraphics[width=2.8in]{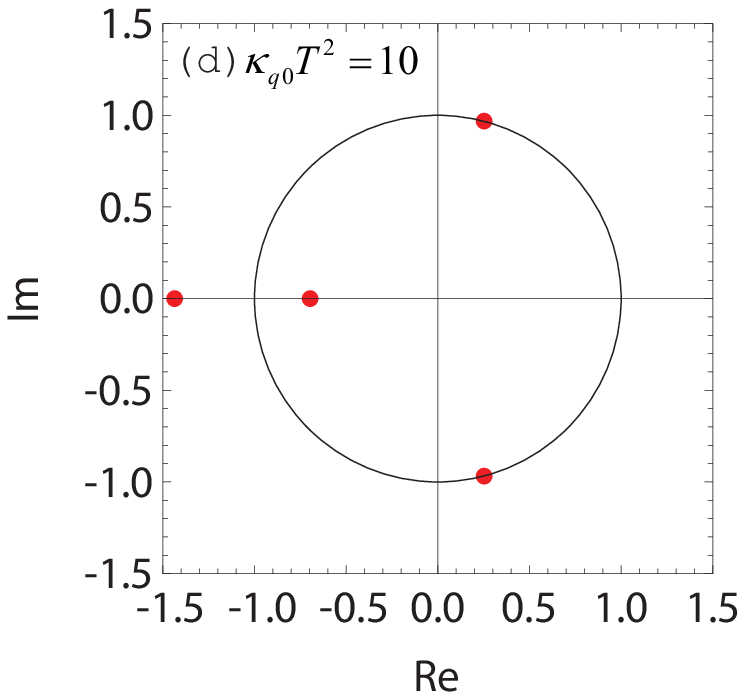} 
\par\end{centering}

\protect\caption{Eigenvalues of continuously rotating quadrupole lattices. The two
eigenvalues on the left half circle have different signatures in (a),
and move towards one other as $\kappa_{q0}T^{2}$ increases in (b).
The Krein collision occurs at $\kappa_{q0}T^{2}=\pi^{2}$ in (c),
after which these two eigenvalues on the left move off the unit circle
and lead to an unstable lattice in (d). }

\label{Rq} 
\end{figure}

As an illustrative application of the theoretical formalism developed
in Secs.\,\ref{sec:GCS} and \ref{sec:Stability-Analysis}, we investigate
here the stability properties of a continuously rotating quadrupole
lattice \cite{Gluckstern79,Roberson1985,Chernin1988,Petillo1989,Krall1995,Smith1998,Petillo2002}.
The Hamiltonian of a charged particle in such a lattice is \cite{Qin11-056708,Chung13}
\begin{align}
H & =\frac{1}{2}z^{T}\Lambda z\,,\,\,\,\Lambda=\left(\begin{array}{cc}
\kappa\left(t\right) & 0\\
0 & I
\end{array}\right)\,,\label{H-1}\\
 & \kappa(t)=\kappa_{q0}\left(\begin{array}{cc}
\cos(2\pi t/T) & \sin(2\pi t/T)\\
-\sin(2\pi t/T) & \cos(2\pi t/T)
\end{array}\right).
\end{align}
We will smoothly vary $\kappa_{q0}T^{2}$ and observe the movement
of the eigenvalues of the one-period map. For a given value of $\kappa_{q0}T^{2}$,
we find a matched $\beta$ solution using the procedure described
above. The calculation is carried out using a code developed in Ref.\,\cite{Chung13}.
The matched $\beta$ solutions for the cases of $\kappa_{q0}T^{2}=8$
and $\kappa_{q0}T^{2}=9$ are plotted in Fig.\,\ref{ms}. The eigenvalue
distributions for different values of $\kappa_{q0}T^{2}$ are plotted
in Fig.\,\ref{Rq}. For $\kappa_{q0}T^{2}=8$ and $\kappa_{q0}T^{2}=9$,
it belongs to Case (1). The two eigenvalues on the left half circle
have different signatures, and move towards one another as $\kappa_{q0}T^{2}$
increases. At $\kappa_{q0}T^{2}=\pi^{2}$, these two eigenvalues collide
at $\lambda=-1.$ Since their signatures are different, this is an
unstable Krein collision and thus structurally unstable. Right after
the collision at $\kappa_{q0}T^{2}=10$, these two eigenvalues moves
off the unit circle, and lead to an unstable lattice. This is the
scenario depicted in Fig.\,\ref{Krein}(b). These analyses can be
straightforwardly carried out for any coupled lattice, such as the
N-rolling lattice \cite{Qin11-056708} and the M\"{o}bius accelerator
\cite{Talman95}. We note that to calculate the eigenvalues displayed
in Fig.\,\ref{Rq}, it is not necessary to find the matched $\beta$
solutions. Any solution of the $\beta$ matrix over one lattice period
can be used. Matched solutions are preferable when the $\beta$ matrix
is used as a lattice function for the machine. 

\section{Conclusions and future work\label{sec:Conclusions}}

We have studied in this paper the spectral and structural stability
of charged particle dynamics in a coupled focusing lattice as a Hamiltonian
system. The recently developed generalized Courant-Synder theory for
coupled lattices has been applied. It is has been demonstrated that
for coupled lattices that are spectrally and structurally stable,
the matrix envelope equation must admit matched solutions. A new method
is presented to determine a matched solution for the matrix envelope
using the technique of normal form and pre-Iwasawa decomposition.
If a matched solution exists, this method is able to determine the
matched solution simply by solving the envelope equation once without
using the (inefficient) shooting procedure. As an example, stability
properties of a continuously rotating quadrupole lattice are investigated.
The Krein collision process for destabilization of the lattice has
also been demonstrated. 

The application of coupled lattices to high-intensity charged particle
beams \cite{Wang82,Chao93-all,Davidson01-284,Lund04,Zimmermann04-124801}
will be investigated in future studies. The theoretical framework
and analytical tools developed in the present study are also expected
to be effective for the current theoretical and experimental investigations
of emittance exchange technologies \cite{Cornacchia02,Emma06,Qin11-758,Groening11,Xiao13-PRSTAB,Groening14,Chung15}.
\begin{acknowledgments}
This research was supported by the U.S. Department of Energy (DE-AC02-09CH11466)
and the 2014 Research Fund(1.140075.01) of UNIST (Ulsan National Institute
of Science and Technology). We thank Prof. Phil Morrison for the interesting
discussion on Krein collision. 
\end{acknowledgments}

\bibliographystyle{apsrev}

\begin{thebibliography}{50}
\expandafter\ifx\csname natexlab\endcsname\relax\def\natexlab#1{#1}\fi
\expandafter\ifx\csname bibnamefont\endcsname\relax
  \def\bibnamefont#1{#1}\fi
\expandafter\ifx\csname bibfnamefont\endcsname\relax
  \def\bibfnamefont#1{#1}\fi
\expandafter\ifx\csname citenamefont\endcsname\relax
  \def\citenamefont#1{#1}\fi
\expandafter\ifx\csname url\endcsname\relax
  \def\url#1{\texttt{#1}}\fi
\expandafter\ifx\csname urlprefix\endcsname\relax\def\urlprefix{URL }\fi
\providecommand{\bibinfo}[2]{#2}
\providecommand{\eprint}[2][]{\url{#2}}

\bibitem[{\citenamefont{Courant and Snyder}(1958)}]{Courant58}
\bibinfo{author}{\bibfnamefont{E.}~\bibnamefont{Courant}} \bibnamefont{and}
  \bibinfo{author}{\bibfnamefont{H.}~\bibnamefont{Snyder}},
  \bibinfo{journal}{Annals of Physics} \textbf{\bibinfo{volume}{3}},
  \bibinfo{pages}{1} (\bibinfo{year}{1958}).

\bibitem[{\citenamefont{Gluckstern}(1979)}]{Gluckstern79}
\bibinfo{author}{\bibfnamefont{R.~L.} \bibnamefont{Gluckstern}}, in
  \emph{\bibinfo{booktitle}{Proceedings of the 1979 Linear Accelerator
  Conference}} (\bibinfo{publisher}{Brookhaven National Laboratory},
  \bibinfo{year}{1979}), pp. \bibinfo{pages}{245--248}.

\bibitem[{\citenamefont{Roberson et~al.}(1985)\citenamefont{Roberson, Mondelli,
  and Chernin}}]{Roberson1985}
\bibinfo{author}{\bibfnamefont{C.}~\bibnamefont{Roberson}},
  \bibinfo{author}{\bibfnamefont{A.}~\bibnamefont{Mondelli}}, \bibnamefont{and}
  \bibinfo{author}{\bibfnamefont{D.}~\bibnamefont{Chernin}},
  \bibinfo{journal}{Particle Accelerators} \textbf{\bibinfo{volume}{17}},
  \bibinfo{pages}{79} (\bibinfo{year}{1985}).

\bibitem[{\citenamefont{Chernin}(1988)}]{Chernin1988}
\bibinfo{author}{\bibfnamefont{D.}~\bibnamefont{Chernin}},
  \bibinfo{journal}{Particle Accelerators} \textbf{\bibinfo{volume}{24}},
  \bibinfo{pages}{29} (\bibinfo{year}{1988}).

\bibitem[{\citenamefont{Petillo et~al.}(1989)\citenamefont{Petillo, Chernin,
  and Mondelli}}]{Petillo1989}
\bibinfo{author}{\bibfnamefont{J.}~\bibnamefont{Petillo}},
  \bibinfo{author}{\bibfnamefont{D.}~\bibnamefont{Chernin}}, \bibnamefont{and}
  \bibinfo{author}{\bibfnamefont{A.}~\bibnamefont{Mondelli}}, in
  \emph{\bibinfo{booktitle}{Proc. of the 1989 Particle Accelerator Conference}}
  (\bibinfo{publisher}{IEEE}, \bibinfo{address}{Piscataway, NJ},
  \bibinfo{year}{1989}), pp. \bibinfo{pages}{1055--1057}.

\bibitem[{\citenamefont{Krall et~al.}(1995)\citenamefont{Krall, Slinker, Lampe,
  and Joyce}}]{Krall1995}
\bibinfo{author}{\bibfnamefont{J.}~\bibnamefont{Krall}},
  \bibinfo{author}{\bibfnamefont{S.}~\bibnamefont{Slinker}},
  \bibinfo{author}{\bibfnamefont{M.}~\bibnamefont{Lampe}}, \bibnamefont{and}
  \bibinfo{author}{\bibfnamefont{G.}~\bibnamefont{Joyce}},
  \bibinfo{journal}{Journal of Applied Physics} \textbf{\bibinfo{volume}{77}},
  \bibinfo{pages}{463} (\bibinfo{year}{1995}).

\bibitem[{\citenamefont{Smith et~al.}(1998)\citenamefont{Smith, Bailey,
  Lackner, and Putnam}}]{Smith1998}
\bibinfo{author}{\bibfnamefont{J.}~\bibnamefont{Smith}},
  \bibinfo{author}{\bibfnamefont{V.}~\bibnamefont{Bailey}},
  \bibinfo{author}{\bibfnamefont{H.}~\bibnamefont{Lackner}}, \bibnamefont{and}
  \bibinfo{author}{\bibfnamefont{S.}~\bibnamefont{Putnam}}, in
  \emph{\bibinfo{booktitle}{Proc. of the 1997 Particle Accelerator Conference}}
  (\bibinfo{publisher}{IEEE}, \bibinfo{address}{Piscataway, NJ},
  \bibinfo{year}{1998}), \bibinfo{number}{1}, pp. \bibinfo{pages}{1251--1253}.

\bibitem[{\citenamefont{Petillo et~al.}(2002)\citenamefont{Petillo, Kostas,
  Chernin, and Mondelli}}]{Petillo2002}
\bibinfo{author}{\bibfnamefont{J.}~\bibnamefont{Petillo}},
  \bibinfo{author}{\bibfnamefont{C.}~\bibnamefont{Kostas}},
  \bibinfo{author}{\bibfnamefont{D.}~\bibnamefont{Chernin}}, \bibnamefont{and}
  \bibinfo{author}{\bibfnamefont{A.}~\bibnamefont{Mondelli}}, in
  \emph{\bibinfo{booktitle}{Proc. of the 1991 Particle Accelerator Conference}}
  (\bibinfo{publisher}{IEEE}, \bibinfo{address}{Piscataway, NJ},
  \bibinfo{year}{2002}), pp. \bibinfo{pages}{613--615}.

\bibitem[{\citenamefont{Talman}(1995)}]{Talman95}
\bibinfo{author}{\bibfnamefont{R.}~\bibnamefont{Talman}},
  \bibinfo{journal}{Physical Review Letters} \textbf{\bibinfo{volume}{74}},
  \bibinfo{pages}{1590} (\bibinfo{year}{1995}).

\bibitem[{\citenamefont{Barnard}(1996)}]{Barnard96}
\bibinfo{author}{\bibfnamefont{J.~J.} \bibnamefont{Barnard}}, in
  \emph{\bibinfo{booktitle}{Proc. of the 1995 Particle Accelerators
  Conference}} (\bibinfo{publisher}{IEEE}, \bibinfo{address}{Piscataway, NJ},
  \bibinfo{year}{1996}), p. \bibinfo{pages}{3241}.

\bibitem[{\citenamefont{Kishek et~al.}(1999)\citenamefont{Kishek, Barnard, and
  Grote}}]{Kishek99}
\bibinfo{author}{\bibfnamefont{R.~A.} \bibnamefont{Kishek}},
  \bibinfo{author}{\bibfnamefont{J.~J.} \bibnamefont{Barnard}},
  \bibnamefont{and} \bibinfo{author}{\bibfnamefont{D.~P.} \bibnamefont{Grote}},
  in \emph{\bibinfo{booktitle}{Proc. of the 1999 Particle Accelerator
  Conference}} (\bibinfo{publisher}{IEEE}, \bibinfo{address}{Piscataway, NJ},
  \bibinfo{year}{1999}), p. \bibinfo{pages}{1761}.

\bibitem[{\citenamefont{Krein}(1950)}]{Krein50}
\bibinfo{author}{\bibfnamefont{M.}~\bibnamefont{Krein}},
  \bibinfo{journal}{Doklady Akad. Nauk. SSSR N.S.}
  \textbf{\bibinfo{volume}{73}}, \bibinfo{pages}{445} (\bibinfo{year}{1950}).

\bibitem[{\citenamefont{Gel'fand and Lidskii}(1955)}]{Gelfand55}
\bibinfo{author}{\bibfnamefont{I.~M.} \bibnamefont{Gel'fand}} \bibnamefont{and}
  \bibinfo{author}{\bibfnamefont{V.~B.} \bibnamefont{Lidskii}},
  \bibinfo{journal}{Uspekhi Mat. Nauk} \textbf{\bibinfo{volume}{10}},
  \bibinfo{pages}{3} (\bibinfo{year}{1955}).

\bibitem[{\citenamefont{Moser}(1958)}]{Moser58}
\bibinfo{author}{\bibfnamefont{J.}~\bibnamefont{Moser}},
  \bibinfo{journal}{Communications on Pure and Applied Mathematics}
  \textbf{\bibinfo{volume}{XI}}, \bibinfo{pages}{81} (\bibinfo{year}{1958}).

\bibitem[{\citenamefont{Yakubovich and Starzhinskii}(1975)}]{Yakubovich75}
\bibinfo{author}{\bibfnamefont{V.}~\bibnamefont{Yakubovich}} \bibnamefont{and}
  \bibinfo{author}{\bibfnamefont{V.}~\bibnamefont{Starzhinskii}},
  \emph{\bibinfo{title}{Linear Differential Equations with Periodic
  Coefficients}}, vol.~\bibinfo{volume}{I} (\bibinfo{publisher}{Wiley},
  \bibinfo{year}{1975}).

\bibitem[{\citenamefont{Dragt}(2014)}]{DragtBook}
\bibinfo{author}{\bibfnamefont{A.~J.} \bibnamefont{Dragt}},
  \emph{\bibinfo{title}{Lie Methods for Nonlinear Dynamics with Applications to
  Accelerator Physics}} (\bibinfo{publisher}{In preparation},
  \bibinfo{year}{2014}).

\bibitem[{\citenamefont{Qin et~al.}(2014)\citenamefont{Qin, Davidson, Burby,
  and Chung}}]{Qin14-044001}
\bibinfo{author}{\bibfnamefont{H.}~\bibnamefont{Qin}},
  \bibinfo{author}{\bibfnamefont{R.~C.} \bibnamefont{Davidson}},
  \bibinfo{author}{\bibfnamefont{J.~W.} \bibnamefont{Burby}}, \bibnamefont{and}
  \bibinfo{author}{\bibfnamefont{M.}~\bibnamefont{Chung}},
  \bibinfo{journal}{Phys. Rev. ST Accel. Beams} \textbf{\bibinfo{volume}{17}},
  \bibinfo{pages}{044001} (\bibinfo{year}{2014}).

\bibitem[{\citenamefont{Teng}(1971)}]{Teng71}
\bibinfo{author}{\bibfnamefont{L.~C.} \bibnamefont{Teng}},
  \bibinfo{journal}{NAL Report FN-229}  (\bibinfo{year}{1971}).

\bibitem[{\citenamefont{Edwards and Teng}(1973)}]{Edwards73}
\bibinfo{author}{\bibfnamefont{D.~A.} \bibnamefont{Edwards}} \bibnamefont{and}
  \bibinfo{author}{\bibfnamefont{L.~C.} \bibnamefont{Teng}},
  \bibinfo{journal}{IEEE Trans. Nucl. Sci.} \textbf{\bibinfo{volume}{NS-20}},
  \bibinfo{pages}{885} (\bibinfo{year}{1973}).

\bibitem[{\citenamefont{Teng}(2003)}]{Teng03}
\bibinfo{author}{\bibfnamefont{L.~C.} \bibnamefont{Teng}}, in
  \emph{\bibinfo{booktitle}{Proceedings of the 2003 Particle Accelerator
  Conference}} (\bibinfo{address}{Piscataway, NJ}, \bibinfo{year}{2003}), p.
  \bibinfo{pages}{2895}.

\bibitem[{\citenamefont{Ripken}(1970)}]{Ripken70}
\bibinfo{author}{\bibfnamefont{G.}~\bibnamefont{Ripken}}, \bibinfo{type}{Tech.
  Rep.} \bibinfo{number}{R1-70/04}, \bibinfo{institution}{DESY}
  (\bibinfo{year}{1970}).

\bibitem[{\citenamefont{Wiedemann}(2007)}]{Wiedemann07-614}
\bibinfo{author}{\bibfnamefont{H.}~\bibnamefont{Wiedemann}},
  \emph{\bibinfo{title}{Particle Accelerator Physics}}
  (\bibinfo{publisher}{Springer-Verlag}, \bibinfo{year}{2007}), pp.
  \bibinfo{pages}{614--620}.

\bibitem[{\citenamefont{Lebedev and Bogacz}(2010)}]{Lebedev10}
\bibinfo{author}{\bibfnamefont{V.~A.} \bibnamefont{Lebedev}} \bibnamefont{and}
  \bibinfo{author}{\bibfnamefont{S.~A.} \bibnamefont{Bogacz}},
  \bibinfo{journal}{Journal of Instrumentation} \textbf{\bibinfo{volume}{5}},
  \bibinfo{pages}{P10010} (\bibinfo{year}{2010}).

\bibitem[{\citenamefont{Dattoli
  et~al.}(1992{\natexlab{a}})\citenamefont{Dattoli, Mari, Richetta, and
  Torre}}]{Dattoli92}
\bibinfo{author}{\bibfnamefont{G.}~\bibnamefont{Dattoli}},
  \bibinfo{author}{\bibfnamefont{C.}~\bibnamefont{Mari}},
  \bibinfo{author}{\bibfnamefont{M.}~\bibnamefont{Richetta}}, \bibnamefont{and}
  \bibinfo{author}{\bibfnamefont{A.}~\bibnamefont{Torre}},
  \bibinfo{journal}{Nuovo Cimento} \textbf{\bibinfo{volume}{107B}},
  \bibinfo{pages}{269} (\bibinfo{year}{1992}{\natexlab{a}}).

\bibitem[{\citenamefont{Dattoli
  et~al.}(1992{\natexlab{b}})\citenamefont{Dattoli, Gallerano, Mari, Torre, and
  Richetta}}]{Dattoli92b}
\bibinfo{author}{\bibfnamefont{G.}~\bibnamefont{Dattoli}},
  \bibinfo{author}{\bibfnamefont{G.}~\bibnamefont{Gallerano}},
  \bibinfo{author}{\bibfnamefont{C.}~\bibnamefont{Mari}},
  \bibinfo{author}{\bibfnamefont{A.}~\bibnamefont{Torre}}, \bibnamefont{and}
  \bibinfo{author}{\bibfnamefont{M.}~\bibnamefont{Richetta}},
  \bibinfo{journal}{Nuovo Cimento} \textbf{\bibinfo{volume}{107B}},
  \bibinfo{pages}{1151} (\bibinfo{year}{1992}{\natexlab{b}}).

\bibitem[{\citenamefont{Dattoli
  et~al.}(1992{\natexlab{c}})\citenamefont{Dattoli, Mari, Mezi, and
  Torre}}]{Dattoli92c}
\bibinfo{author}{\bibfnamefont{G.}~\bibnamefont{Dattoli}},
  \bibinfo{author}{\bibfnamefont{C.}~\bibnamefont{Mari}},
  \bibinfo{author}{\bibfnamefont{L.}~\bibnamefont{Mezi}}, \bibnamefont{and}
  \bibinfo{author}{\bibfnamefont{A.}~\bibnamefont{Torre}},
  \bibinfo{journal}{Nucl. Instr. Methods Phys. Res.}
  \textbf{\bibinfo{volume}{A321}}, \bibinfo{pages}{447}
  (\bibinfo{year}{1992}{\natexlab{c}}).

\bibitem[{\citenamefont{Qin and Davidson}(2009{\natexlab{a}})}]{Qin09-NA}
\bibinfo{author}{\bibfnamefont{H.}~\bibnamefont{Qin}} \bibnamefont{and}
  \bibinfo{author}{\bibfnamefont{R.~C.} \bibnamefont{Davidson}},
  \bibinfo{journal}{Physical Review Special Topics - Accelerators and Beams}
  \textbf{\bibinfo{volume}{12}}, \bibinfo{pages}{064001}
  (\bibinfo{year}{2009}{\natexlab{a}}).

\bibitem[{\citenamefont{Qin and Davidson}(2009{\natexlab{b}})}]{Qin09PoP-NA}
\bibinfo{author}{\bibfnamefont{H.}~\bibnamefont{Qin}} \bibnamefont{and}
  \bibinfo{author}{\bibfnamefont{R.~C.} \bibnamefont{Davidson}},
  \bibinfo{journal}{Phys. Plasmas} \textbf{\bibinfo{volume}{16}},
  \bibinfo{pages}{050705} (\bibinfo{year}{2009}{\natexlab{b}}).

\bibitem[{\citenamefont{Qin et~al.}(2009)\citenamefont{Qin, Chung, and
  Davidson}}]{Qin09-PRL}
\bibinfo{author}{\bibfnamefont{H.}~\bibnamefont{Qin}},
  \bibinfo{author}{\bibfnamefont{M.}~\bibnamefont{Chung}}, \bibnamefont{and}
  \bibinfo{author}{\bibfnamefont{R.~C.} \bibnamefont{Davidson}},
  \bibinfo{journal}{Physical Review Letters} \textbf{\bibinfo{volume}{103}},
  \bibinfo{pages}{224802} (\bibinfo{year}{2009}).

\bibitem[{\citenamefont{Chung et~al.}(2010)\citenamefont{Chung, Qin, and
  Davidson}}]{Chung10}
\bibinfo{author}{\bibfnamefont{M.}~\bibnamefont{Chung}},
  \bibinfo{author}{\bibfnamefont{H.}~\bibnamefont{Qin}}, \bibnamefont{and}
  \bibinfo{author}{\bibfnamefont{R.~C.} \bibnamefont{Davidson}},
  \bibinfo{journal}{Physics of Plasmas} \textbf{\bibinfo{volume}{17}},
  \bibinfo{pages}{084502} (\bibinfo{year}{2010}).

\bibitem[{\citenamefont{Qin and Davidson}(2011)}]{Qin11-056708}
\bibinfo{author}{\bibfnamefont{H.}~\bibnamefont{Qin}} \bibnamefont{and}
  \bibinfo{author}{\bibfnamefont{R.~C.} \bibnamefont{Davidson}},
  \bibinfo{journal}{Physics of Plasmas} \textbf{\bibinfo{volume}{18}},
  \bibinfo{pages}{056708} (\bibinfo{year}{2011}).

\bibitem[{\citenamefont{Qin et~al.}(2013)\citenamefont{Qin, Davidson, Chung,
  and Burby}}]{Qin13PRL2}
\bibinfo{author}{\bibfnamefont{H.}~\bibnamefont{Qin}},
  \bibinfo{author}{\bibfnamefont{R.~C.} \bibnamefont{Davidson}},
  \bibinfo{author}{\bibfnamefont{M.}~\bibnamefont{Chung}}, \bibnamefont{and}
  \bibinfo{author}{\bibfnamefont{J.~W.} \bibnamefont{Burby}},
  \bibinfo{journal}{Phy. Rev. Lett.} \textbf{\bibinfo{volume}{111}},
  \bibinfo{pages}{104801} (\bibinfo{year}{2013}).

\bibitem[{\citenamefont{Qin and Davidson}(2013)}]{Qin13PRL}
\bibinfo{author}{\bibfnamefont{H.}~\bibnamefont{Qin}} \bibnamefont{and}
  \bibinfo{author}{\bibfnamefont{R.~C.} \bibnamefont{Davidson}},
  \bibinfo{journal}{Physical Review Letters} \textbf{\bibinfo{volume}{110}},
  \bibinfo{pages}{064803} (\bibinfo{year}{2013}).

\bibitem[{\citenamefont{Chung et~al.}(2013)\citenamefont{Chung, Qin, Gilson,
  and Davidson}}]{Chung13}
\bibinfo{author}{\bibfnamefont{M.}~\bibnamefont{Chung}},
  \bibinfo{author}{\bibfnamefont{H.}~\bibnamefont{Qin}},
  \bibinfo{author}{\bibfnamefont{E.~P.} \bibnamefont{Gilson}},
  \bibnamefont{and} \bibinfo{author}{\bibfnamefont{R.~C.}
  \bibnamefont{Davidson}}, \bibinfo{journal}{Phys. Plasmas}
  \textbf{\bibinfo{volume}{20}}, \bibinfo{pages}{083121}
  (\bibinfo{year}{2013}).

\bibitem[{\citenamefont{Iwasawa}(1949)}]{Iwasawa49}
\bibinfo{author}{\bibfnamefont{K.}~\bibnamefont{Iwasawa}},
  \bibinfo{journal}{Annals of Mathematics} \textbf{\bibinfo{volume}{50}},
  \bibinfo{pages}{507} (\bibinfo{year}{1949}).

\bibitem[{\citenamefont{de~Gosson}(2006)}]{deGosson06-42}
\bibinfo{author}{\bibfnamefont{M.}~\bibnamefont{de~Gosson}},
  \emph{\bibinfo{title}{Symplectic Geometry and Quantum Mechanics}}
  (\bibinfo{publisher}{Birkhauser Verlag}, \bibinfo{year}{2006}), pp.
  \bibinfo{pages}{42--44}.

\bibitem[{\citenamefont{Michelotti}(1995)}]{Micheloetti95-166}
\bibinfo{author}{\bibfnamefont{L.}~\bibnamefont{Michelotti}},
  \emph{\bibinfo{title}{Intermediate classical dynamics with applications to
  beam physics}} (\bibinfo{publisher}{John Wiley \& Sons Inc.},
  \bibinfo{year}{1995}), p. \bibinfo{pages}{166}.

\bibitem[{\citenamefont{Hoffstaetter}(1995)}]{Hoffstaetter95}
\bibinfo{author}{\bibfnamefont{G.~H.} \bibnamefont{Hoffstaetter}}, in
  \emph{\bibinfo{booktitle}{Proceedings of 1995 Parcticle Accelerator
  Conference}} (\bibinfo{year}{1995}), pp. \bibinfo{pages}{2707--2710}.

\bibitem[{\citenamefont{Wang and Smith}(1982)}]{Wang82}
\bibinfo{author}{\bibfnamefont{T.~F.} \bibnamefont{Wang}} \bibnamefont{and}
  \bibinfo{author}{\bibfnamefont{L.}~\bibnamefont{Smith}},
  \bibinfo{journal}{Part. Accel.} \textbf{\bibinfo{volume}{12}},
  \bibinfo{pages}{247} (\bibinfo{year}{1982}).

\bibitem[{\citenamefont{Chao}(1993)}]{Chao93-all}
\bibinfo{author}{\bibfnamefont{A.~W.} \bibnamefont{Chao}},
  \emph{\bibinfo{title}{Physics of Collective Beam Instabilities in High Energy
  Accelerators}} (\bibinfo{publisher}{Wiley}, \bibinfo{address}{NewYork},
  \bibinfo{year}{1993}).

\bibitem[{\citenamefont{Davidson and Qin}(2001)}]{Davidson01-284}
\bibinfo{author}{\bibfnamefont{R.~C.} \bibnamefont{Davidson}} \bibnamefont{and}
  \bibinfo{author}{\bibfnamefont{H.}~\bibnamefont{Qin}},
  \emph{\bibinfo{title}{Physics of Intense Charged Particle Beams in High
  Energy Accelerators}} (\bibinfo{publisher}{World Scientific},
  \bibinfo{address}{Singapore}, \bibinfo{year}{2001}), p. \bibinfo{pages}{284}.

\bibitem[{\citenamefont{Lund and Bukh}(2004)}]{Lund04}
\bibinfo{author}{\bibfnamefont{S.}~\bibnamefont{Lund}} \bibnamefont{and}
  \bibinfo{author}{\bibfnamefont{B.}~\bibnamefont{Bukh}},
  \bibinfo{journal}{Physical Review Special Topics - Accelerators and Beams}
  \textbf{\bibinfo{volume}{7}}, \bibinfo{pages}{024801} (\bibinfo{year}{2004}).

\bibitem[{\citenamefont{Zimmermann}(2004)}]{Zimmermann04-124801}
\bibinfo{author}{\bibfnamefont{F.}~\bibnamefont{Zimmermann}},
  \bibinfo{journal}{Physical Review Special Topic - Accelerators and Beams}
  \textbf{\bibinfo{volume}{7}}, \bibinfo{pages}{124801} (\bibinfo{year}{2004}).

\bibitem[{\citenamefont{Cornacchia and Emma}(2002)}]{Cornacchia02}
\bibinfo{author}{\bibfnamefont{M.}~\bibnamefont{Cornacchia}} \bibnamefont{and}
  \bibinfo{author}{\bibfnamefont{P.}~\bibnamefont{Emma}},
  \bibinfo{journal}{Physical Review Special Topics - Accelerators and Beams}
  \textbf{\bibinfo{volume}{5}}, \bibinfo{pages}{1} (\bibinfo{year}{2002}).

\bibitem[{\citenamefont{Emma et~al.}(2006)\citenamefont{Emma, Huang, Kim, and
  Piot}}]{Emma06}
\bibinfo{author}{\bibfnamefont{P.}~\bibnamefont{Emma}},
  \bibinfo{author}{\bibfnamefont{Z.}~\bibnamefont{Huang}},
  \bibinfo{author}{\bibfnamefont{K.-J.} \bibnamefont{Kim}}, \bibnamefont{and}
  \bibinfo{author}{\bibfnamefont{P.}~\bibnamefont{Piot}},
  \bibinfo{journal}{Physical Review Special Topics - Accelerators and Beams}
  \textbf{\bibinfo{volume}{9}}, \bibinfo{pages}{100702} (\bibinfo{year}{2006}).

\bibitem[{\citenamefont{Qin et~al.}(2011)\citenamefont{Qin, Davidson, Chung,
  Barnard, and Wang}}]{Qin11-758}
\bibinfo{author}{\bibfnamefont{H.}~\bibnamefont{Qin}},
  \bibinfo{author}{\bibfnamefont{R.~C.} \bibnamefont{Davidson}},
  \bibinfo{author}{\bibfnamefont{M.}~\bibnamefont{Chung}},
  \bibinfo{author}{\bibfnamefont{J.~J.} \bibnamefont{Barnard}},
  \bibnamefont{and} \bibinfo{author}{\bibfnamefont{T.~F.} \bibnamefont{Wang}},
  in \emph{\bibinfo{booktitle}{Proceedings of 2011 Particle Accelerator
  Conference}} (\bibinfo{publisher}{IEEE}, \bibinfo{address}{Piscataway, NJ},
  \bibinfo{year}{2011}), p. \bibinfo{pages}{758}.

\bibitem[{\citenamefont{Groening}(2011)}]{Groening11}
\bibinfo{author}{\bibfnamefont{L.}~\bibnamefont{Groening}},
  \bibinfo{journal}{Phys. Rev. ST Accel. Beams} \textbf{\bibinfo{volume}{14}},
  \bibinfo{pages}{064201} (\bibinfo{year}{2011}).

\bibitem[{\citenamefont{Xiao et~al.}(2013)\citenamefont{Xiao, Kester, Groening,
  Leibrock, Maier, and Rottl\"ander}}]{Xiao13-PRSTAB}
\bibinfo{author}{\bibfnamefont{C.}~\bibnamefont{Xiao}},
  \bibinfo{author}{\bibfnamefont{O.}~\bibnamefont{Kester}},
  \bibinfo{author}{\bibfnamefont{L.}~\bibnamefont{Groening}},
  \bibinfo{author}{\bibfnamefont{H.}~\bibnamefont{Leibrock}},
  \bibinfo{author}{\bibfnamefont{M.}~\bibnamefont{Maier}}, \bibnamefont{and}
  \bibinfo{author}{\bibfnamefont{P.}~\bibnamefont{Rottl\"ander}},
  \bibinfo{journal}{Phys. Rev. ST Accel. Beams} \textbf{\bibinfo{volume}{16}},
  \bibinfo{pages}{044201} (\bibinfo{year}{2013}).

\bibitem[{\citenamefont{Groening et~al.}(2014)\citenamefont{Groening, Maier,
  Xiao, Dahl, Gerhard, Kester, Mickat, Vormann, Vossberg, and
  Chung}}]{Groening14}
\bibinfo{author}{\bibfnamefont{L.}~\bibnamefont{Groening}},
  \bibinfo{author}{\bibfnamefont{M.}~\bibnamefont{Maier}},
  \bibinfo{author}{\bibfnamefont{C.}~\bibnamefont{Xiao}},
  \bibinfo{author}{\bibfnamefont{L.}~\bibnamefont{Dahl}},
  \bibinfo{author}{\bibfnamefont{P.}~\bibnamefont{Gerhard}},
  \bibinfo{author}{\bibfnamefont{O.~K.} \bibnamefont{Kester}},
  \bibinfo{author}{\bibfnamefont{S.}~\bibnamefont{Mickat}},
  \bibinfo{author}{\bibfnamefont{H.}~\bibnamefont{Vormann}},
  \bibinfo{author}{\bibfnamefont{M.}~\bibnamefont{Vossberg}}, \bibnamefont{and}
  \bibinfo{author}{\bibfnamefont{M.}~\bibnamefont{Chung}},
  \bibinfo{journal}{Physical Review Letters} \textbf{\bibinfo{volume}{113}},
  \bibinfo{pages}{264802} (\bibinfo{year}{2014}).

\bibitem[{\citenamefont{Chung et~al.}(2015)\citenamefont{Chung, Qin, Groening,
  C.Davidson, and Xiao}}]{Chung15}
\bibinfo{author}{\bibfnamefont{M.}~\bibnamefont{Chung}},
  \bibinfo{author}{\bibfnamefont{H.}~\bibnamefont{Qin}},
  \bibinfo{author}{\bibfnamefont{L.}~\bibnamefont{Groening}},
  \bibinfo{author}{\bibfnamefont{R.}~\bibnamefont{C.Davidson}},
  \bibnamefont{and} \bibinfo{author}{\bibfnamefont{C.}~\bibnamefont{Xiao}},
  \bibinfo{journal}{Physics of Plasmas} \textbf{\bibinfo{volume}{22}},
  \bibinfo{pages}{013109} (\bibinfo{year}{2015}).

\end{thebibliography}

\end{document}